\documentclass[12pt,preprint]{aastex}
\usepackage{amsmath}
\usepackage{mathrsfs}
\usepackage{latexsym,amsfonts,amssymb}
\usepackage{natbib}
\usepackage{graphicx}
\begin{document}
\title{{ Orbit}-induced spin precession as an origin of periodicity in periodically-repeating fast radio bursts  }

\author{Huan Yang\altaffilmark{1,2}, Yuan-Chuan Zou\altaffilmark{1,3}}

\altaffiltext{1}{Perimeter Institute for Theoretical Physics, Waterloo, Ontario N2L 2Y5, Canada. Email: hyang@perimeterinstitute.ca (HY)}
\altaffiltext{2}{University of Guelph, Guelph, Ontario N2L 3G1, Canada}
\altaffiltext{3}{School of Physics, Huazhong University of Science and
Technology, Wuhan, 430074, China. Email: zouyc@hust.edu.cn (YCZ)}

\begin{abstract}
FRB 180916.J0158+65 has been found to repeatedly emit fast radio bursts with the period in roughly 16 days. We propose that such periodicity comes from { orbit}-induced, spin precession of the emitter, which is possibly a neutron star. Depending on the mass of the companion, the binary period ranges from several hundreds to thousands of seconds. Such tight binaries have a relatively short lifetime, and they are not likely the products of gravitational decay from wide binaries. We comment on the relation to GW190425 and the possibility in LISA and LIGO detections.
\end{abstract}


\maketitle

\section{Introduction}
Fast radio bursts (FRB) are energetic pulses of GHz radio emission
with durations from microseconds to milliseconds
\citep{2007Sci...318..777L, 2013Sci...341...53T, 2015Natur.528..523M, 2016MPLA...3130013K, 2016Sci...354.1249R, 2019ApJ...885L..24C}. They have dispersion measures (DMs) between 300-1500 pc cm$^{-3}$, which is much
larger than the line of sight DM contribution expected from the
electron distribution of our Galaxy. Later on, the observation of the host galaxy of FRB 121102 confirmed the extra-Galactic origin of FRBs \citep{2017Natur.541...58C,2017ApJ...834L...7T}, which is also the first repeating FRB \citep{2016Natur.531..202S}. Despite the unclear nature of FRBs, the confirmation of cosmological origin makes FRBs as the cosmological probes, such as constraining the baryon number density \citep{2014ApJ...783L..35D, 2016Natur.530..453K, 2019MNRAS.484.1637J},
measuring cosmic proper distance \citep{2017A&A...606A...3Y}, finding missing
baryons \citep{2014ApJ...780L..33M}, constraining dark energy
\citep{2014PhRvD..89j7303Z, 2014ApJ...788..189G},  and testing Einstein's Equivalent Principle
\citep{2015PhRvL.115z1101W}.

Up to now, more than 1000 FRBs have been detected \citep{2016PASA...33...45P}, of which are mostly contributed by the Canadian Hydrogen Intensity Mapping Experiment Fast Radio Burst Project (CHIME/FRB) \citep{2018ApJ...863...48C, 2019Natur.566..230C}.
FRB 180916.J0158+65 was first detected by CHIME together with the other 8 new repeating FRBs \citep{2019ApJ...885L..24C}. With more than one year of operation, CHIME found a $16.35\pm 0.18$  day periodicity \citep{2020arXiv200110275T}. This is the first FRB with periodicity identified.

Many models have been proposed to explain the extraordinary features of FRBs \citep[for recent reviews,
see][]{2016MPLA...3130013K,2019PhR...821....1P}.
Among all different models, to explain the repeating FRBs, the neutron star is mostly engaged \footnote{See however the shortcomings of neutron star model, and an explanation with black hole jets \citep{2019arXiv191200526K}.}, such as flares of the magnetars \citep{2010vaoa.conf..129P, 2014ApJ...797...70K, 2014MNRAS.442L...9L, 2017ApJ...843L..26B, 2019MNRAS.485.4091M}, similar origin to soft gamma-ray repeaters \citep{2016ApJ...826..226K}, giant pulses from young pulsars\citep{2016MNRAS.462..941L}, curvature radiation from the strong magnetic field of neutron stars \citep{2017MNRAS.468.2726K}, interaction of inspiraling double neutron stars \citep{2016ApJ...822L...7W, 2020arXiv200200335Z}, and even may arise the connection of FRBs, gamma-ray bursts and gravitational wave bursts \citep{2014ApJ...780L..21Z, 2016ApJ...827L..31Z}.
Based on the single neutron star model producing the fast radio emission, here we concentrate on the possible origin of the $16.35\pm 0.18$  day periodicity. { We suggest the period could be arisen from orbit-induced, spin precession of a neutron star. This kind of precession has been suggested from modelling long term polarized position angle of 81 pulsars by \citet{2010ApJ...721.1044W}, with precession period in the range of 200-1300 days.}
We show the procession model in section \ref{sec:180916}, and the conclusion and discussion are presented in section \ref{sec:discussion}.

\section{Spin precession of FRB 180916.J0158+65} \label{sec:180916}

\begin{figure}\label{fig:sketch}
    \centering
    \includegraphics[height=0.3\textheight]{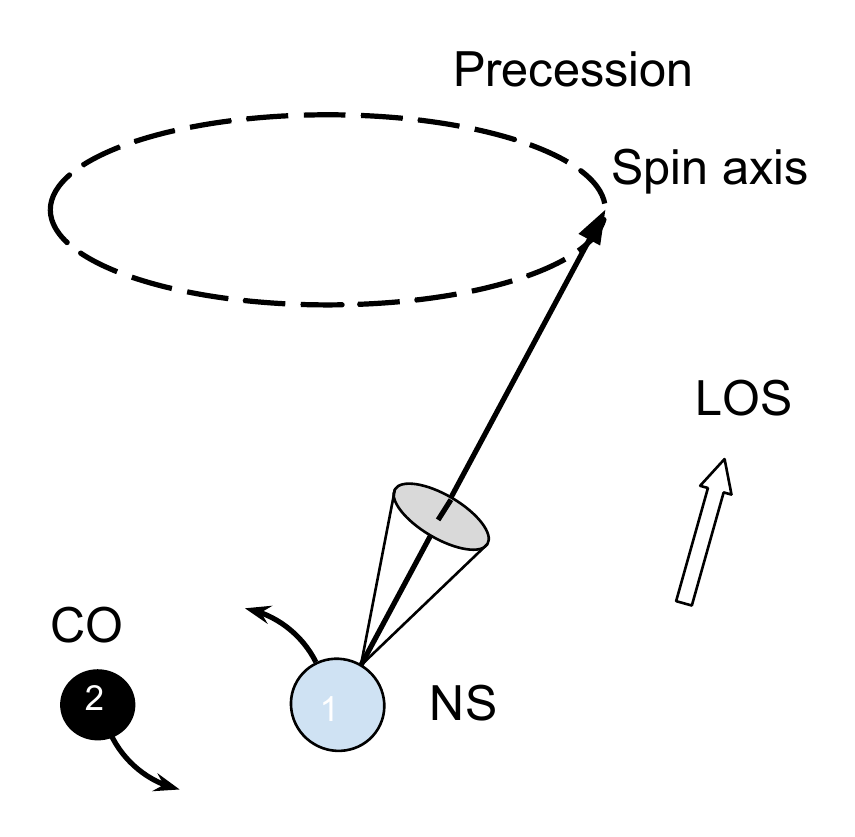}
    \includegraphics[height=0.3\textheight]{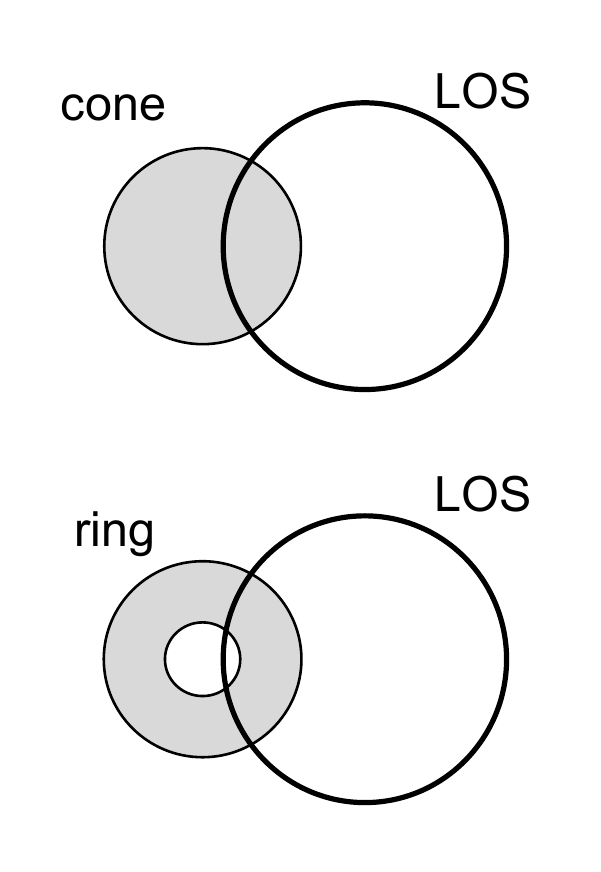}
    \caption{The left panel shows the schematic plot of the model. The neutron star (NS) emits sparse FRBs inside a solid angle along the spin axis. The emitting region is painted in grey, which is much larger than the opening angle of each individual FRB. If the emission is similar to a pulsar, the region could be a ring rather than a polar cap. The companion, indicating as { CO (Compact Object), which could be a stellar mass Black Hole, a Neutron Star or a White Dwarf}. The precession is shown as a dashed circle. The line of sight (LOS) is plotted as an arrow. The right panel shows the configuration of the emitting region relating to the line of sight in the frame of the neutron star, where the precession axis locates at the center of the circles. The overlap of the emitting region and the line of sight shows the observable fraction. The upper right sub-panel corresponds to the case if the FRB emitting region locates in the whole region along the spin axis. The lower right sub-panel corresponds to the case that the emitting cone rotates along the spin axis, which is similar to a pulsar emitting ring.}
\end{figure}

Let us consider a compact binary system with mass $M_1$ and $M_2 = q M_1$ for the emitter and its companion respectively. The emitter, which is likely a neutron star, may rotate around its spin axis with periods range from milliseconds (millisecond pulsars) to seconds (magnetars). Therefore the emission pattern in the emitter's sky may look like a disk or a ring depending on the opening angle of the emission with respect to the spin axis and the width of the emission cone. A schematic plot is given in Fig. \ref{fig:sketch}.

The spin ${ S_1}$ of the emitter may also precess due to spin-obit coupling \citep{2014grav.book.....P}:
\begin{align}\label{eq:prec}
\frac{d { S}_1}{d t} =& \frac{G}{2a^3{ (1-e^2)^{3/2}}c^2}\left [\left ( 4+\frac{3 M_2}{M_1} \right ) { L}\right ] \times { S}_1\,,
\end{align}
{ where $e$ is the orbit eccentricity, $G$ is the gravitational constant, $c$ is the speed of light}, ${ L}$ is the the orbital angular momentum vector and $a$ is the separation between two objects. We have neglected possible spin-spin coupling between the FRB emitter and its companion, which may become important if the companion is an intermediate mass black hole. With Eq.~\ref{eq:prec}, the spin precession frequency is just $\Omega_{\rm prec} =(4+3 q) (G L/c^2) /( 2a^3 ({ \sqrt{1-e^2}})^3)$ which we shall identify as $16$ days for FRB 180916.J0158+65. The orbital frequency is given by
\begin{align}\label{eq:o}
\Omega_{\rm orb} =\sqrt{\frac{G (M_1+M_2)}{a^3}} =\sqrt{\frac{G M_1(1+q)}{a^3}}\,,
\end{align}
and the magnitude of orbital angular momentum is
\begin{align}\label{eq:l}
L = \frac{M_1 M_2}{(M_1+M_2)} \Omega_{\rm orb} a^2
{ \sqrt{1-e^2}}\,.
\end{align}

\begin{figure}
\includegraphics[scale=0.85]{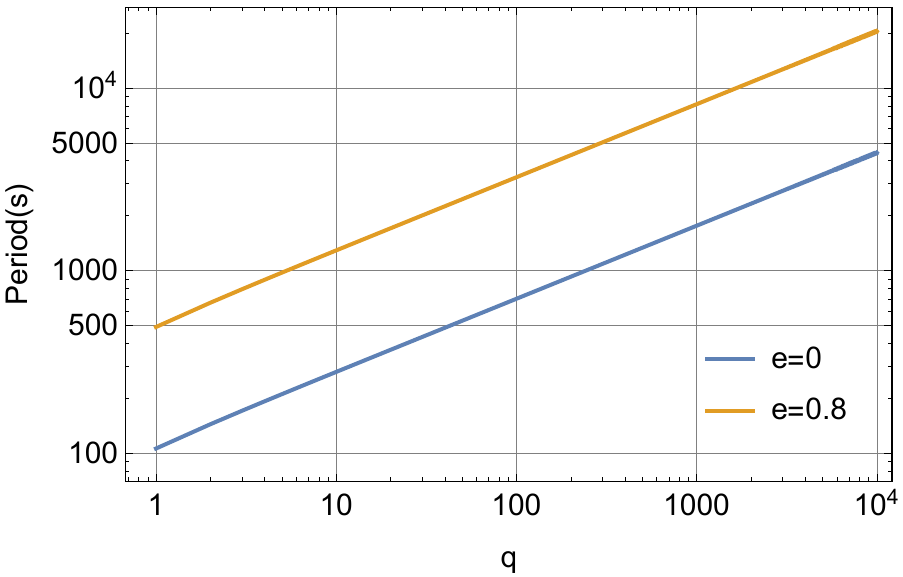}
\includegraphics[scale=0.85]{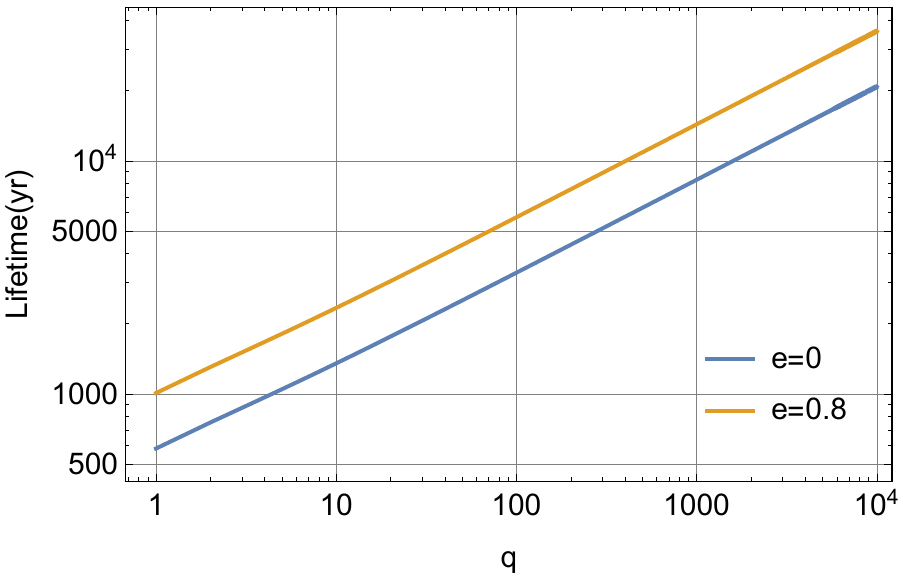}
\caption{\label{fig:period} Left Panel: the period of the compact binary as a function of the binary mass ratio. The emitter is assumed to be a 1.4 $M_\odot$ neutron star, and its spin precession period is 16 days. Right panel: The lifetime of such binaries under gravitational wave radiation.}
\end{figure}

\begin{figure}
\includegraphics[scale=0.65]{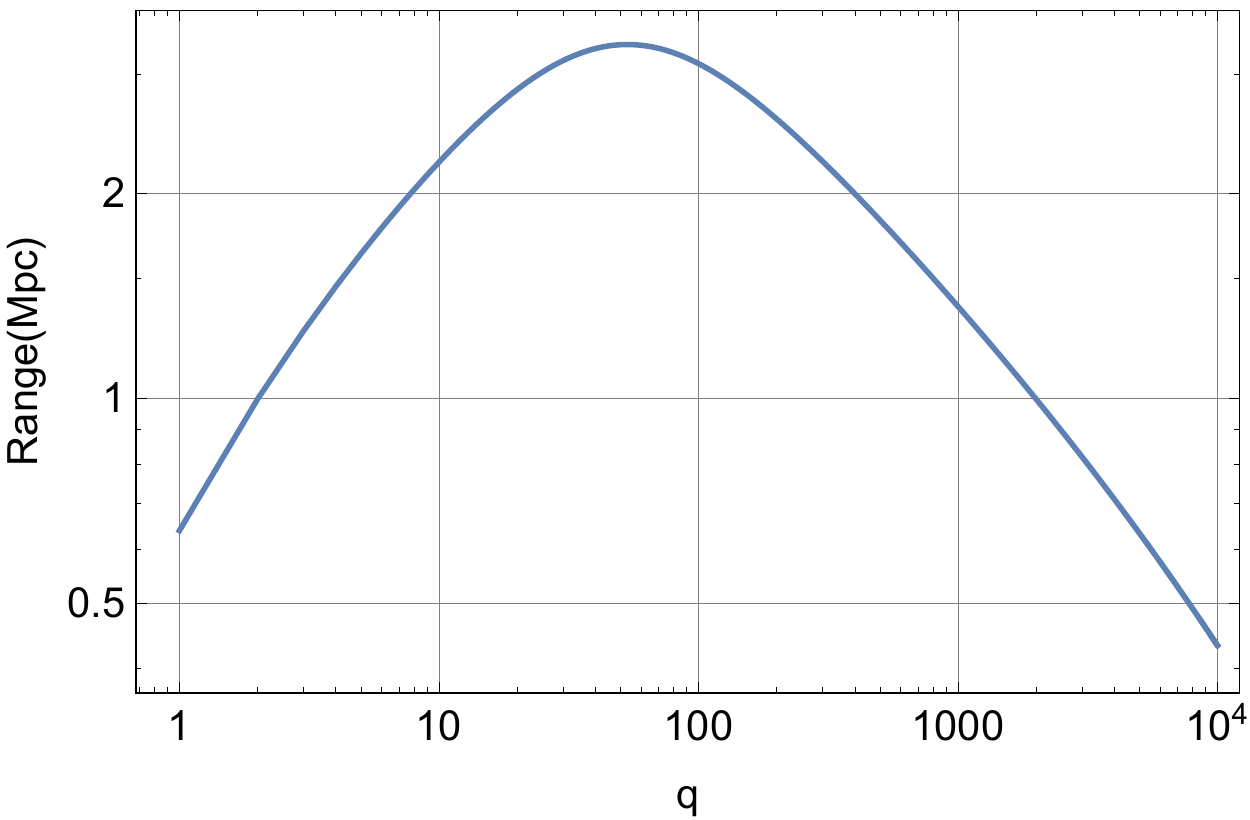}
\caption{\label{fig:range} The LISA detection range of systems considered in Fig.~\ref{fig:period}, with detection threshold SNR set to be 15. { The orbit is assumed to be circular for simplicity.}}
\end{figure}

\begin{figure}
\includegraphics[scale=0.65]{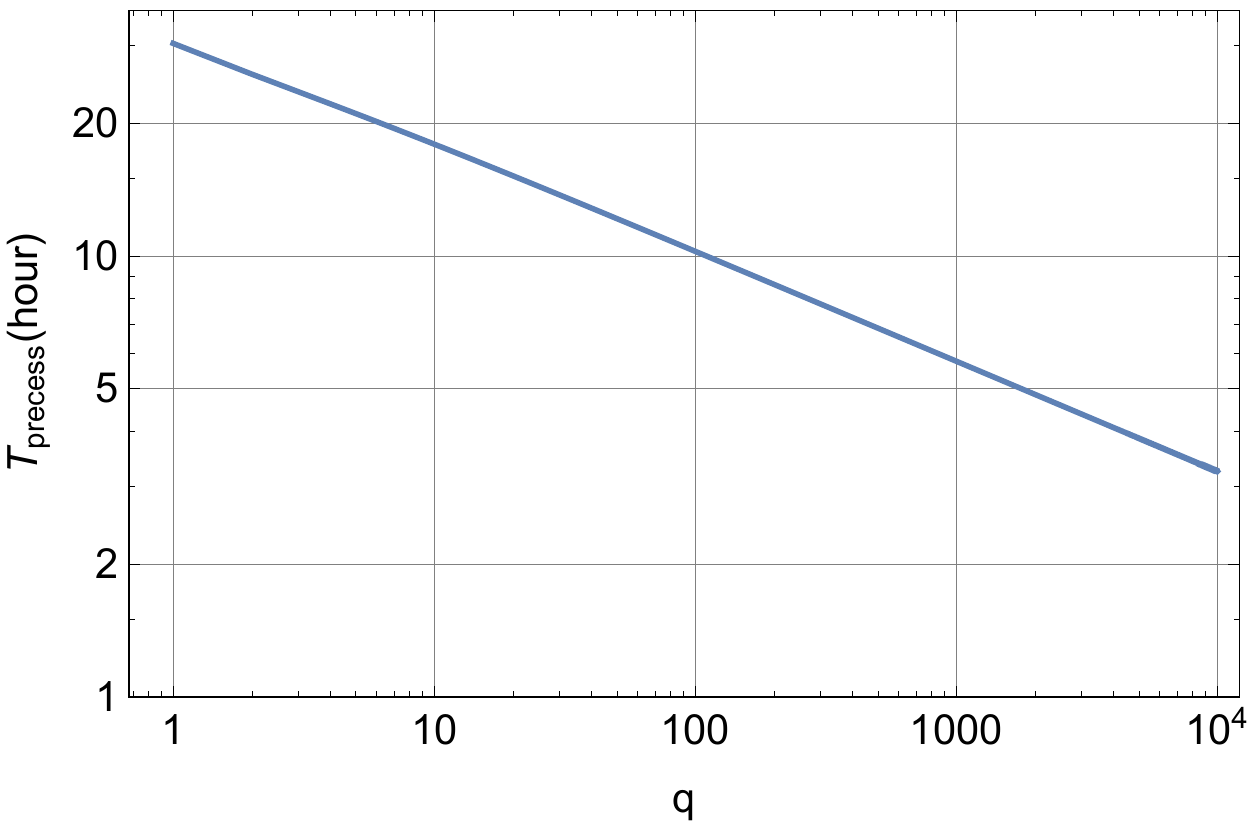}
\caption{\label{fig:precess} The expected precession timescale of the $1.4 M_\odot$ neutron star within a stellar mass binary, assuming that the binary lifetime is ten years and { the orbit is circular}. This setting is useful for searching periodic FRBs that have the potential of multi-messenger detection. }
\end{figure}

Combining Eq.~\ref{eq:prec}, Eq.~\ref{eq:o} and Eq.~\ref{eq:l}, we can easily obtain the period of the binary as a function of the mass ratio, as depicted in Fig.~\ref{fig:period}. It is also well approximated as
\begin{align}
T \approx \frac{10^2}{ (1-e^2)^{3/2}} q^{0.4} {\rm s}\,.
\end{align}

For stellar-mass binaries and binaries including an intermediate mass black hole, the orbital periods range from several hundreds of seconds to several thousands of seconds. As the FRB emitter is away from the galaxy center, it is unlikely that the companion is a supermassive black hole.
The inferred distance is much smaller than the radius of a normal star, so that the companion has to be a compact object. Such periods are smaller than those of any known neutron star binaries in our galaxy \footnote{The shortest orbital period is in order of 100 minutes as shown at https://www.atnf.csiro.au/research/pulsar/psrcat/}. In addition, the lifetime of such a system is \citep{Peters:1964zz}
\begin{align}
T_{\rm GW} = \frac{12}{19} \frac{c^4_0}{\beta} \int^e_0 d e \frac{x^{29/19} [1+(121/304)x^2]^{1181/2299}}{(1-x^2)^{3/2}}\,,
\end{align}
with
\begin{align}
c_0 =a (1-e^2) e^{-12/19} \left [ 1+\frac{121}{304}e^2\right ]^{-870/2299},\quad \beta =\frac{64}{5} \frac{G^3 M_1 M_2 (M_1+M_2)}{c^5}
\end{align}
as shown in the right panel of Fig.~\ref{fig:period}. These timescales are order of thousands of years for stellar-mass binaries, which are much shorter than the typical lifetimes of field binaries \footnote{In fact, the lifetimes of double neutron star systems observed in the range from 86 Myr to well beyond Hubble time \citep{Tauris:2017omb}.}.
As $T_{\rm GW} \propto a^4$ and $T_{\rm prec} \propto a^{5/2}$, it is reasonable to expect that the distribution of similar systems in precession time
should be
\begin{align}
\frac{dN}{d T_{\rm prec}} \propto T_{\rm prec}^{3/5}\,,
\end{align}
unless there is a cutoff in the corresponding lifetime. This feature can be used to test this model as we find more repeating FRBs with periodicity. If there is an upper cutoff in precession time, we can use it to find the initial distance of the binary when it is formed.
For FRB 180916.J0158+65, if the emitter is a magnetar with an active time $\sim 10^4$ years, this system is unlikely originated from gravitational wave decay from a typical, wide field binary.
In addition, if we have indeed observed a short-period system, we should have observed much more FRBs (more than 1000 of FRB detection so far) given the large ratio between wide binaries and such tight binary.
It may instead comes from formation channels with much closer distances.

Such possibility may be related to the heavy binary neutron star mergers observed in GW190425 \citep{Abbott:2020uma}, which is heavier than any known binary neutron stars in our galaxy \footnote{It is still possible that one (or both)  of the compact objects in the binary system of GW 190425 is a black hole \citep{Yang:2017gfb,Abbott:2020uma,Han:2020qmn,Kyutoku:2020xka}.}.  It was suggested that tight binaries  exist in our galaxy, because short-period binaries are difficult to observe with current methods.
On the other hand, if this heavy neutron star pair is formed within a fast-merging channel, this may also account for the potential bias towards detection \citep{Abbott:2020uma,Romero-Shaw:2020aaj}.
 \cite{Safarzadeh:2020efa} argued that this leads to theoretical difficulties as traditional fast-merging channels (such as case BB unstable mass transfer) are unable to produce as frequent neutron star mergers. Nevertheless, these new observations will shed light on the formation processes of neutron stars.


{ These} tight binaries fall into the detection band of LISA (Laser Interferometer Space Antenna) \citep{2000AdSpR..25.1129D}, or similar space-borne gravitational wave detector such as Tianqin \citep {2016CQGra..33c5010L} or Taiji \citep{10.1093/nsr/nwx116}. For quasi-circular source, the event SNR can be computed as
\begin{align}
{\rm SNR} \approx 2 |h(f)|\sqrt{\frac{\Delta f}{S_{\rm LISA}(f)}}\,,
\end{align}
where $S_{\rm LISA}$ is the detector spectral density, $h(f)$ is the frequency domain waveform, $\Delta f= \dot{f} T_{\rm obs}$ is the frequency shift during the observation period and $\dot{f}$ is induced by gravitational wave radiation. Setting the detection threshold for the signal to noise ratio to be 15, we plot the range of detection for LISA in Fig.~\ref{fig:range}, with the observation time $T_{\rm obs}$ taken as 5 years. { We have assumed circular orbits here to avoid dealing with multi-frequency emission from elliptical orbits.} FRB 180916.J0158+65 is estimated to be $\sim 100$ Mpc away, which is clearly outside the detection range. However, if the FRB emission is beamed, it is reasonable to imagine that there are closer sources with FRB emission not pointing towards us. If the solid angle of emission is $\Delta \Omega$, then for any source observed with FRB emission at distance d, it is reasonable to expect another source as close as $d (\Delta \Omega/4\pi)^{1/3}$. If we take into account that maybe a fraction $\eta$ of these sources are FRB emitters (i.e., the fraction of the active phase), then the estimator becomes
\begin{align}
d_{\rm min} = d \left(\eta  \frac{\Delta \Omega}{4\pi} \right)^{1/3} \sim 1.3 {\rm Mpc} \left ( \frac{d}{100 {\rm Mpc}}\right ) \left ( \frac{\Delta \Omega}{10 {\rm deg}^2}\right )^{1/3}  \left ( \frac{\eta}{10^{-2}}\right )^{1/3}\,.
\end{align}
 In any case, it seems to be challenging to have a multi-messenger observation of such periodic repeaters with both FRB and gravitational wave measurements.

 On the other hand, suppose many of such periodic FRBs are indeed stellar-mass binaries  in tight orbits. It is instructive to imagine as we find more of such systems, especially the ones at larger distances, some of them may have a lifetime on the order of ten years instead of thousands of years. As a result, we may first identify such periodic repeaters through FRB observation, and then measure the gravitational wave signals associated with binary mergers later on.  Notice that the detection range of Advanced LIGO (Laser Interferometric Gravitational Wave Observatory) is a few hundred Mpcs and for Advanced LIGO Plus it is roughly $z \sim 0.2$ \citep{2019BAAS...51c.141R,barsotti2018a+} \footnote{Notice that if the companion is a white dwarf, the binary will merge before entering LIGO band, although the merging process may produce strong electromagnetic emissions.}. Suppose the lifetime is ten years, we can work out the expected precession timescales, which are several hours, as shown in Fig.~\ref{fig:precess}. Therefore it may be useful to search for periodic repeaters with such periodicity in the future.
This proposal itself can serve as a check for this precession model of periodic FRB repeaters.
If we indeed observe such short-precession time binary, we may even monitor the time-dependent evolution of the precession period, which should be $T_{\rm prec} \propto (t_c-t)^{5/8}$, with $t_c$ being the coalescence time.

\section{Conclusion and Discussion} \label{sec:discussion}

In this work, we examine the possibility that FRB 180916.J0158+65 comes from a compact binary system, while the spin precession of FRB emitter gives rise to the periodicity observed. This interpretation naturally predicts the period of the binary, which ranges from a few hundred seconds to a few thousand seconds, depending on the binary mass ratio. Such type of tight binaries has recently been discussed for GW190425 in order to account for possible observational bias of short-period systems.
Because of the relatively short lifetime, we expect this system also comes from a fast-merging channel, instead of being the product of gravitational wave decay from a wide binary. Besides the argument based on binary lifetime and rates, in light of the LIGO detection of heavy binary neutron stars in GW190425, it is also interesting to recognize that the more possible type of FRB emitters, magnetars, indeed are more massive than normal  pulsars as they may come from more massive progenitors \citep{Gaensler_2005}.

Although the period of such systems fit directly into the LISA observation band ($10^{-4}$-$10^{-1}$ Hz), it will be difficult to find them detachable  in FRBs and gravitational waves.
It is also unclear whether it is possible to extract the orbital period by  electromagnetic observations, as FRBs repeat rather infrequently.
However, it might be possible to find periodic repeaters with periods as several hours, which may lead to opportunities of gravitational wave detection by LIGO/Virgo or their upgrades later on. In addition, if the FRB emitter is a magnetar \footnote{Magnetars are thought to ``last" for $\sim 10^4$ years as the magnetic field gradually decays. This could be longer than the lifetime of such binary.} and its companion is a black hole with a mass ratio $\ge 5$, the merger product will be a temporarily charged black hole. The discharge process takes several milliseconds associated with strong gamma ray emission, which may be observed by Fermi-LAT at $\sim 100$Mpc distance (assuming $B \sim 10^{15}$ G) or a Cherenkov telescope such as \cite{Ong:2019zyq} \citep{Pan:2019ulx}.

We have included eccentricity in this simple model of orbit. For binaries produced by case BB unstable mass transfer, the eccentricity may already be small \citep{Romero-Shaw:2020aaj}, depending on the magnitude of supernova kicks. For binaries produced by dynamical interactions, the associated eccentricities have a small chance to  be significant ($e > 0.1$) \citep{Rodriguez:2018pss,Kremer:2018cir,Pan:2019uyn}.

During the preparation of the manuscript, \cite{2020arXiv200201920L} propose a model that a pulsar in tight early B-star binary can arise the 16 days periodicity due to orbital phase-dependent modulation. \cite{2020arXiv200201920L} disfavoured the geodetic precession scenario { by considering the energy budget from the orbit-induced electric potential. This is different from the model studied here, where we assume that the emission comes from a single emitter instead of the interaction between two stellar objects. Recently \cite{Levin:2020rhj} and \cite{Zanazzi:2020vyp} proposed to use a deformed precessing magnetar to explain the observed periodicity. Notice that in their models the precession rate should decay with time as the magnetar rotation slows down or the star deformation relaxes, whereas in this orbit-induced precession scenario the precession rate actually increases with time as the orbit shrinks by gravitational wave radiation.}


 We thank Zhen Pan for helping with the fitting function of LISA sensitivity curve, and Jonathan Katz for comments. H. Y. is supported by the
Natural Sciences and Engineering Research Council of
Canada and in part by Perimeter Institute for Theoretical
Physics. Research at Perimeter Institute is supported
in part by the Government of Canada through the Department
of Innovation, Science and Economic Development
Canada and by the Province of Ontario through the
Ministry of Economic Development, Job Creation and
Trade. Y.-C. Z is supported by the National Natural Science Foundation of China (Grants No. U1738132, 11773010 and U1931203).

\bibliography{frb}

\end{document}